\documentclass[twocolumn,pra,aps,showpacs]{revtex4}

\usepackage{psfrag,graphicx}

\usepackage{ae,aecompl}

\usepackage{overpic}

\usepackage{epsfig}

\usepackage{dcolumn}

\usepackage{amsmath,amssymb}

\usepackage{bm}

\usepackage[dvips]{color}

\definecolor{mygrey}{gray}{0.35}

\definecolor{mygreen}{rgb}{0.85,1,0.9}

\definecolor{myzard}{cmyk}{0,0,0.05,0}

\definecolor{mywhite}{rgb}{1,1,1}

\definecolor{myred}{rgb}{1,0,0}

 \def\ee{\mathord{\rm e}}

 \def\ii{\mathord{\rm i}}

\def\half{\textstyle\frac{1}{2}}

\def\vec#1{{\bf{#1}}}

\def\bra#1{\langle#1|} \def\ket#1{|#1\rangle}

\begin{document}

\title[Short Title]{
Chirality Quantum Phase Transition in the Dirac oscillator}

\author{A. Bermudez$^1$, M.A. Martin-Delgado$^1$ and A.Luis$^{2}$}

\affiliation{ $^1$Departamento de F\'{\i}sica Te\'orica I,
Universidad Complutense, 28040 Madrid, Spain \\$^2$Departamento de
 \'Optica, Universidad Complutense, 28040 Madrid,
Spain}

\begin{abstract}

We study a relativistic spin-1/2 fermion subjected to a Dirac
oscillator coupling and a constant magnetic field. An interplay
between opposed chirality interactions culminates in the
appearance of a relativistic quantum phase transition, which can
be fully characterized. We obtain analytical expressions for the
energy gap, order parameter, and canonical quantum fluctuations
across the critical point. Moreover, we also discuss the effect of
this phase transition on the statistics of the chiral bosonic
ensemble, where its super- or sub-Poissonian nature can be
controled by means of external parameters. Finally, we study the
entanglement properties between the degrees of freedom in the
relativistic ground state, where an interesting transition between
a bi-separable and a genuinely tripartite entangled state occurs.

\end{abstract}


\maketitle

\section{Introduction}

Phase transitions describe an abrupt change in the physical
properties of a system caused by the modification of its
temperature. This phenomenon usually entails a change in the
symmetry of the phases involved, and is commonly driven by thermal
fluctuations. Consequently, classical phase transition cannot
persist at zero temperature where the absence of thermal
fluctuations forbids the sudden change of phase. Nonetheless,
other kind of fluctuations exist at zero temperature, the so
called quantum fluctuations, which can also be responsible for a
dramatic change in the properties of the system. In this case, the
change is driven by the modification of certain couplings that
describe the interactions between the microscopic constituents of
the system, and is usually known as a quantum phase
transition~\cite{Sachved}.

These critical phenomena arise in the thermodynamical limit of
certain many-body systems, where the number of particles
$N\to\infty$. Usually, the description of such systems is
extremely complex and one must deal with numerical methods.
Nevertheless, there exist certain situations where a simplified
model, which can be exactly solved, captures the full essence of
the problem displaying such an abrupt change of the system
properties. For instance, a collection of $N$ two-level atoms
interacting with a single mode of the radiation field, known as
the Dicke model, displays a quantum phase transition whose
features can be justified by means of a simple two-mode
Hamiltonian in the thermodynamical
limit~\cite{Dicke_pt,Dicke_chaos}. Other two-mode Hamiltonians
which also display critical phenomena have been studied in the
field of Quantum Optics~\cite{gerry_pt, gerry_pt_II}, or in
Nuclear Physics~\cite{Jahn_teller_millburn}.

In this work, we shall be concerned with a relativistic toy model
that involves two phonon modes and also displays a quantum phase
transition. This critical phenomenon occurs in a fermionic
relativistic harmonic oscillator, also known as a Dirac
oscillator~\cite{moshinsky, ito}, when an additional constant
magnetic field is applied. This relativistic fermion of mass $m$
and charge $-e$ is described by the following Dirac equation
\begin{equation}
\label{dirac_oscillator_magnetic_field_3D}
 \ii\hbar\frac{\partial |\Psi\rangle}{\partial
t}=\left[c\boldsymbol{\alpha}\left(\textbf{p}-\ii m\beta\omega
\textbf{r}+e\textbf{A}\right)+\beta mc^2\right]|\Psi\rangle,
\end{equation}
where $\ket{\Psi}$ stands for the Dirac 4-component spinor,
$\textbf{r}$ and $\textbf{p}$ represent the position and momentum
operators, $\omega$ is the Dirac oscillator frequency, $c$ stands
for the speed of light and
$\beta:=\text{diag}(\mathbb{I},-\mathbb{I}),\alpha_j:=\text{off-diag}(\sigma_j,\sigma_j)$
are the Dirac matrices related to the usual Pauli matrices
\cite{greiner}. Here, the magnetic field is introduced by minimal
coupling $ \textbf{p}\to\textbf{p}+e\vec{A}$ , where $\vec{A}$ is
the vector potential related to the magnetic field through
$\vec{B}=\nabla\wedge\vec{A}$. On the other hand, the Dirac
oscillator coupling is introduced by a non-minimal coupling
$\textbf{p}-\ii m\beta\omega \textbf{r}$, where $\omega$ stands
for the Dirac oscillator frequency.

Here, we shall focus on a two-dimensional setup where the Dirac
matrices become the well-known Pauli matrices:
$\alpha_x=\sigma_x,\alpha_y=\sigma_y,\beta=\sigma_z$. In this
scenario, Eq.~\eqref{dirac_oscillator_magnetic_field_3D} can be
expressed as follows
\begin{equation}
\label{dirac_oscillator_magnetic_field_2D}
 \ii\hbar\frac{\partial |\Psi\rangle}{\partial
t}=\left[c\sum_{j=1}^2\sigma_j\left(p_j-\ii m\beta\sigma_z
x_j+eA_j\right)+\sigma_z mc^2\right]|\Psi\rangle,
\end{equation}
where $|\Psi\rangle$ is a 2-component spinor which mixes spin-up
and -down components with positive and negative energies.
Remarkably enough, the Dirac oscillator coupling endows the
particle with an intrinsic left-handed chirality which is only
present in this two-dimensional scenario~\cite{bermudez_do_2D}.
Conversely, the magnetic field coupling favors a right-handed
chirality~\cite{bermudez_dirac_cats}, and therefore an intriguing
interplay is set up. This system can be considered as a
relativistic extension of chiral harmonic oscillators, which carry
dual aspects of a certain symmetry (i.e. chirality in the plane).
Such non-relativistic systems have been studied from a fundamental
point of view~\cite{chiral_banerjee}, and an interesting
connection to topological Chern-Simons gauge theories has also
been pointed out~\cite{chiral_wotza,chern_jackiw}.

In this paper, we show how this relativistic chiral oscillator
presents several intriguing  critical properties, and offer an
ideal scenario where to study the effect of opposed chirality
interactions in a two-dimensional setting. In
Sec.~\ref{sectionII}, we show how the relativistic Hamiltonian in
Eq.~\eqref{dirac_oscillator_magnetic_field_2D} can be exactly
mapped onto a pair of simultaneous Jaynes-Cummings(JC) and
Anti-Jaynes-Cummings(AJC)
interactions~\cite{jaynes_cummings,wineland_review} with right-
and left-handed chirality respectively. This result differs
substantially from previous situations where only a distinctive
chiral interaction
appeared~\cite{bermudez_do_2D,bermudez_dirac_cats}, and opens up
the possibility to study a unique interplay between left- and
right-handed interactions. In Sec.~\ref{sectionIII}, we describe
two limiting cases: a weak magnetic field regime where the system
displays a remarkable left-handed chirality, and a strong magnetic
field regime with an opposite right-handed chirality. For
intermediate couplings, we obtain the complete energy spectrum and
the associated eigenstates  in Sec.~\ref{sectionIV}. To accomplish
such task, we perform a unitary transformation which turns the
Hamiltonian into a single-mode interaction which can be easily
diagonalized. This exact solution shows how an unusual competition
between chiral terms arises, and leads to a relativistic level
crossing phenomenon for a critical value of the JC and AJC
couplings, which is described in detail in Sec.~\ref{sectionV}. In
this sense, several analogies with second order quantum phase
transitions occur, such as the energy gap suppression,  the
divergence of quantum fluctuations, and the existence of an order
parameter that reveals the existence of a quantum phase
transition. Furthermore, we also show that the statistical nature
of the chiral phonon distribution displays an abrupt change across
the critical point, where super-Poissonian chiral phonons turn
into sub-Poissonian phonons and vice versa. Finally, in
Appendix~\ref{apendix_a}, we give some details for the
construction of the system eigenstates, which are closely related
to $SU(1,1)$ coherent states.

\section{Exact mapping onto a simultaneous JC-AJC Hamiltonian}
\label{sectionII}

\begin{figure}[!hbp]

\centering

\begin{overpic}[width=6.0cm]{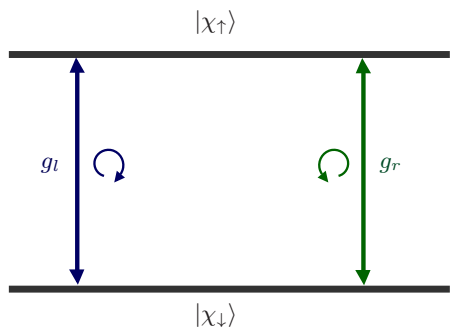}
\put(8,42){\textcolor[rgb]{0.00,0.00,0.33}{$g_l$}}
\put(83,42){\textcolor[rgb]{0.00,0.29,0.15}{$g_r$}}
\put(42,8){$\ket{\chi_{\downarrow}}$}
\put(42,73){$\ket{\chi_{\uparrow}}$}
\end{overpic}
\caption{Level scheme representation of relativistic interaction,
which gives rise to spin-flip transitions accompanied by the
simultaneous creation and annihilation of chiral phonons. Note
that there are two different spin-flip channels depending on the
phonon chirality .}\label{niveles}

\end{figure}

Let us first provide the exact mapping onto a pair of simultaneous
JC and AJC couplings with opposite chiralities. We shall work in
the axial gauge, where a constant magnetic field
$\textbf{B}=B\textbf{e}_z$ is described by the following vector
potential  $\vec{A}:=\frac{B}{2}[-y,x,0]$. In this setup, the
dynamics of a relativistic fermion is described by two different
frequencies, the Dirac oscillator frequency $\omega$ and the
cyclotron frequency $\omega_c:=eB/m$. In this regard, we must
introduce a pair of sets of creation-annihilation operators
\begin{equation}
\label{creation_annihilation_ops}
\begin{split}
&a_i=\frac{1}{\sqrt{2}}\left(\frac{1}{\Delta}r_i + \ii
\frac{\Delta}{\hbar}p_i\right),\hspace{2ex}a^{\dagger}_i=\frac{1}{\sqrt{2}}\left(\frac{1}{\Delta}r_i
- \ii \frac{\Delta}{\hbar}p_i\right),\\
&\tilde{a}_i=\frac{1}{\sqrt{2}}\left(\frac{1}{\tilde{\Delta}}r_i +
\ii
\frac{\tilde{\Delta}}{\hbar}p_i\right),\hspace{2ex}\tilde{a}^{\dagger}_i=\frac{1}{\sqrt{2}}\left(\frac{1}{\tilde{\Delta}}r_i
- \ii \frac{\tilde{\Delta}}{\hbar}p_i\right),\\
\end{split}
\end{equation}
associated to the frequencies $\omega$ and
$\tilde{\omega}:=\omega_c/2$ respectively.  Here,
$\Delta:=\sqrt{\hbar/m\omega}$ and
$\tilde{\Delta}:=\sqrt{\hbar/m\tilde{\omega}}$ represent the
oscillator's ground state width associated to each frequency, and
we have introduced $i=x,y$ to account for the two possible
directions of motion~\cite{notaI}. The substitution of these
operators in Eq.~\eqref{dirac_oscillator_magnetic_field_2D},
followed by the introduction of the chiral annihilation operators
for each frequency
\begin{equation}
\label{circular_operators}
\begin{array}{c}
  \tilde{a}_r:=\frac{1}{\sqrt{2}}(\tilde{a}_x - \ii \tilde{a}_y),\hspace{2ex} \tilde{a}_l:=\frac{1}{\sqrt{2}}(\tilde{a}_x + \ii \tilde{a}_y) , \\
  a_r:=\frac{1}{\sqrt{2}}(a_x - \ii a_y),\hspace{2ex}a_l:=\frac{1}{\sqrt{2}}(a_x + \ii a_y) , \\
\end{array}
\end{equation}
and the consequent creation operators
$\tilde{a}^{\dagger}_r,\tilde{a}^{\dagger}_l,a^{\dagger}_r,a^{\dagger}_r$,
leads to the following bichromatic relativistic Hamiltonian
\begin{equation}
\label{matrix_hamiltonian}
H=mc^2\left[%
\begin{array}{cc}
  1 & \ii \sqrt{2\xi}a^{\dagger}_l-\ii\sqrt{2\tilde{\xi}}\tilde{a}_r  \\
  -\ii \sqrt{2\xi}a_l+\ii\sqrt{2\tilde{\xi}}\tilde{a}^{\dagger}_r & -1 \\
\end{array}%
\right],
\end{equation}
where  the two important parameters $\xi:=\hbar\omega/mc^2$, and
$\tilde{\xi}:=\hbar\tilde{\omega}/mc^2$ represent the strength of
the oscillator and magnetic field coupling with respect to the
particle rest mass energy, respectively. This Hamiltonian in
Eq.~\eqref{matrix_hamiltonian} can be rewritten as a simultaneous
JC and AJC interactions with opposite chiralities
\begin{equation}
\label{JC-AJC-hamiltonian}
H=\delta\sigma_z-H_{\text{JC}}^{\circlearrowleft}(g_r)+H_{\text{AJC}}^{\circlearrowright}(g_l),
\end{equation}
where $\delta:=mc^2$ stands for the detuning parameter
proportional to the rest mass energy,
$H_{\text{JC}}^{\circlearrowleft}(g_r)$ represents a right-handed
Jaynes-Cummings Hamiltonian
\begin{equation}
\label{JC_right}
H_{\text{JC}}^{\circlearrowleft}(g_r)=g_{r}\sigma^{+}\tilde{a}_r+g^*_{r}\sigma^{-}\tilde{a}^{\dagger}_r,
\end{equation}
 with $g_{r}:=\ii mc^2\sqrt{2\tilde{\xi}}$ as the
interaction coupling strength. Analogously, the term
$H_{\text{AJC}}^{\circlearrowright}(g_l)$ stands for a left-handed
 Anti-Jaynes-Cummings interaction
\begin{equation}
\label{AJC_left}
H_{\text{AJC}}^{\circlearrowright}(g_l)=g_{l}\sigma^{+}a^{\dagger}_l+g^*_{l}\sigma^{-}a_l,
\end{equation}
with a similar coupling strength $g_{l}:=\ii mc^2\sqrt{2\xi}$. The
relativistic Hamiltonian in Eq.~\eqref{JC-AJC-hamiltonian} is
depicted in fig.~\ref{niveles}, where the JC and AJC couplings
give rise to a pair of channels through which the relativistic
particle can perform spin-flip transitions.

\section{Weak and strong magnetic field regimes}
\label{sectionIII}

The exact mapping of this relativistic model onto a simultaneous
JC and AJC couplings allows a neat description of the interaction
in terms of a energy level scheme (see fig.~\ref{niveles}). The
presence of the pair of simultaneous couplings
$H_{\text{AJC}}^{\circlearrowright}(g_l)$, and
$H_{\text{JC}}^{\circlearrowleft}(g_r)$ in
Eq.~\eqref{JC-AJC-hamiltonian}, forbids a straightforward solution
of the complete Hamiltonian by analogy to JC-like
solutions~\cite{bermudez_do_2D,bermudez_dirac_cats}. Nonetheless,
these techniques can be used to obtain the energy spectrum and
corresponding eigenstates in two limiting regimes:

\subsection{ Weak magnetic field $\tilde{\xi}\ll\xi$}

In this regime, the Hamiltonian in Eq.~\eqref{matrix_hamiltonian}
becomes
\begin{equation}
H^{w}=2\delta\sigma_z+g_{l}\sigma^{+}a^{\dagger}_l+g^*_{l}\sigma^{-}a_l+\mathcal{O}\left(\tilde{\xi}/\xi\right),
\end{equation}
and we can obtain a first approximation to the energy spectrum
neglecting corrections of order $\mathcal{O}(\tilde{\xi}/\xi)$.
Consequently, the relativistic Hamiltonian can be approximately
described by a left-handed AJC coupling which can be directly
solved. The energy spectrum becomes $E^w=\pm E^w_{n_l}=\pm mc^2[
1+2\xi(n_l+1)]^{1/2}, $ where $n_l=0,1...$ represents the number
of left-handed phonons. The corresponding eigenstates form an
AJC-doublet which can be described as follows
\begin{equation}
\label{left_handed_eigenstates} \ket{\pm
E^w_{n_l}}=C^w_{n_l,\pm}\ket{n_l+1}\ket{\chi_{\uparrow}}\mp\ii
C^w_{n_l,\mp}\ket{n_l}\ket{\chi_{\downarrow}},
\end{equation}
where we have introduced the set of left-handed Fock states
$\ket{n_l}:=(n_l!)^{-1/2}(a_l^\dagger)^{n_l}|\text{vac}\rangle$ ,
and certain normalization constants $C^w_{n_l,\pm}:=[(E^w_{n_l}\pm
mc^2)/2E^w_{n_l}]^{1/2}$.

 \subsection{Strong magnetic field $\tilde{\xi}\gg\xi$}

In this limit, the Hamiltonian in Eq.~\eqref{matrix_hamiltonian}
becomes
\begin{equation}
H^s=2\delta\sigma_z-g_{r}\sigma^{+}\tilde{a}_r-g^*_{r}\sigma^{-}\tilde{a}^{\dagger}_r+\mathcal{O}\left(\xi/\tilde{\xi}\right),
\end{equation}
and we can find the energy spectrum up to
$\mathcal{O}(\xi/\tilde{\xi})$, which becomes $E^s=\pm
E^s_{\tilde{n}_r}=\pm mc^2[ 1+2\tilde{\xi}(\tilde{n}_r+1)]^{1/2}$,
where $\tilde{n}_r=0,1...$ is the number of right-handed phonons.
The corresponding eigenstates form a JC-doublet with the following
structure
\begin{equation}
\label{right_handed_eigenstates} \ket{\pm
E^s_{\tilde{n}_r}}=C^s_{\tilde{n}_r,\pm}\ket{\tilde{n}_r}\ket{\chi_{\uparrow}}\mp\ii
C^s_{\tilde{n}_r,\mp}\ket{\tilde{n}_r+1}\ket{\chi_{\downarrow}},
\end{equation}
with
$\ket{\tilde{n}_r}:=(\tilde{n}_r!)^{-1/2}(\tilde{a}_r^\dagger)^{\tilde{n}_r}|\text{vac}\rangle$
as right-handed Fock states, and
$C^s_{\tilde{n}_r,\pm}:=[(E^s_{\tilde{n}_r}\pm
mc^2)/2E^s_{\tilde{n}_r}]^{1/2}$.

The latter results can be understood as follows, in the regime
where $\tilde{\xi}/\xi\to 0$, the relativistic system is found in
a phase with left chirality. Conversely, in the limit where
$\tilde{\xi}/\xi\to \infty$, a notorious right-handed chiral phase
arises. Therefore, a modification of the coupling strengths
$\xi,\tilde{\xi}$ leads to a change of the chiral symmetry, which
can be interpreted in the language of quantum phase transitions.
We will consider a zero-temperature setup, where thermal
fluctuations do not exist and  the chirality transition can only
be driven by quantum fluctuations. As we show below, the complete
relativistic energy spectrum can be obtained for all possible
couplings $\xi,\tilde{\xi}$. This spectrum presents two crucial
properties which resemble the usual setting in quantum phase
transitions: the energy spectrum is non-analytical for a critical
coupling $(\tilde{\xi}/\xi)_c=1$, and the system becomes gapless
at this critical point $\Delta E \to 0$ when
$|(\tilde{\xi}/\xi)-(\tilde{\xi}/\xi)_c|\to 0$.

\section{Exact solution: Energy spectrum and associated eigenstates}
\label{sectionIV}

 The particular structure of the Hamiltonian in Eq.~\eqref{matrix_hamiltonian} suggests
 a different description where the spinorial degrees of freedom
 are coupled to a  unique bosonic operator $b\propto
 (a_l-(\tilde{\xi}/\xi)^{1/2}\tilde{a}_r^{\dagger})$ which mixes both chiralities depending
 on the relative coupling strength $\tilde{\xi}/\xi$. In this regard, we derive a unitary
transformation which converts the bichromatic Hamiltonian in
Eq.~\eqref{JC-AJC-hamiltonian} into a monochromatic JC (AJC) term
that involves such bosonic degree of freedom $b$ with a certain
chirality that depends on external parameters $\xi>\tilde{\xi}$
($\xi<\tilde{\xi}$). This transformation, which turns out to be a
two-mode squeezing operator, allows an insightful derivation of
the energy spectrum and its associated eigenstates whose
properties depend strongly on the magnitude of the magnetic field
applied. As we shall discuss, certain properties of the system,
such as chirality, squeezing, phonon statistics, and entanglement,
are conditioned to the value of the magnetic field. Moreover, as
the magnetic field is varied, a sudden quantum phase transition
occurs at $\xi=\tilde{\xi}$, where both chiralities contribute
identically, and the critical theory  becomes that of a
relativistic free fermion.

\subsection{Left-handed regime $\tilde{\xi}<\xi$}

 Under
these circumstances, the relativistic Hamiltonian in
Eqs.~\eqref{JC-AJC-hamiltonian}-\eqref{AJC_left} can be unitarily
mapped onto a single-mode Anti-Jaynes-Cummings Hamiltonian by
means of the following unitary transformation
\begin{equation}
\label{unitary_transformation_left} \text{U}_{\alpha}:=
  \ee^{\alpha\left(a_l\tilde{a}_r-\tilde{a}^{\dagger}_ra_l^{\dagger}\right)},
\end{equation}
where the real parameter $\alpha$ depends on the relative strength
of the oscillator and magnetic field couplings
\begin{equation}
\alpha:=\frac{1}{\lambda}\text{arctanh}\left(\frac{\lambda\sqrt{\tilde{\omega}}}{\sqrt{\omega}-\mu\sqrt{\tilde{\omega}}}\right),
\end{equation}
and we have introduced
$\mu:=(\Delta/\tilde{\Delta}-\tilde{\Delta}/\Delta)/2$, and
$\lambda:=\sqrt{\mu^2+1}$. The transformation laws for the chiral
operators under the unitary  in
Eq.~\eqref{unitary_transformation_left} can be described as
follows
\begin{equation}
\label{operators_transformation}
\begin{split}
&\text{U}^{\dagger}_{\alpha}a_l
\text{U}_{\alpha}=\left(\text{cosh}(\lambda\alpha)+\frac{\mu}{\lambda}\text{sinh}(\lambda\alpha)\right)a_l-\frac{1}{\lambda}\text{sinh}(\lambda\alpha)\tilde{a}_r^{\dagger},\\
&\text{U}^{\dagger}_{\alpha}\tilde{a}_r\text{U}_{\alpha}=\left(\text{cosh}(\lambda\alpha)-\frac{\mu}{\lambda}\text{sinh}(\lambda\alpha)\right)\tilde{a}_r-\frac{1}{\lambda}\text{sinh}(\lambda\alpha)a_l^{\dagger},
\end{split}
\end{equation}
which lead to the following transformation of the complete
original Hamiltonian into a single-mode AJC term
\begin{equation}
\label{transformed_AJC_left}
H_{\text{AJC}}^{\circlearrowright}(g'_l):=\text{U}_{\alpha}H\text{U}_{\alpha}^{\dagger}=\delta\sigma_z+g_l'\sigma^{+}a_l^{\dagger}+(g_l')^{*}\sigma^{-}a_l,
\end{equation}
with a modified coupling strength $g'_l:=\ii mc^2\sqrt{2\zeta_l}$,
where
$\zeta_l(\xi,\tilde{\xi}):=\xi-\tilde{\xi}-2\mu(\tilde{\xi}\xi)^{1/2}$
is related to the initial relevant parameters. Once the
single-mode Hamiltonian in Eq.~\eqref{transformed_AJC_left} has
been obtained, we can calculate the energy spectrum and
corresponding eigenstates of the original Hamiltonian by solving
the single mode coupling. Two Hermitian operators related through
a unitary transformation share a common spectrum, and therefore
the energy eigenvalues of the  Hamiltonian in
Eq.~\eqref{JC-AJC-hamiltonian} can be obtained in analogy to the
weak magnetic field regime
\begin{equation}
\label{transformed_left_handed_energy} E=\pm E_{n_l}=\pm
mc^2\sqrt{1+2\zeta_l(\xi,\tilde{\xi})(n_l+1)},
\end{equation}
where $n_l=0,1...$ represents the number of left-handed quanta.
The associated eigenstates are obtained by means of the unitary
transform applied to the single-mode left-handed eigenstates
\begin{equation}
\label{left_squeezed_eigenstates} \ket{\pm
E_{n_l}}=\text{U}_{\alpha}^{\dagger}\left(C_{n_l,\pm}\ket{n_l+1}\ket{\chi_{\uparrow}}\mp\ii
C_{n_l,\mp}\ket{n_l}\ket{\chi_{\downarrow}}\right),
\end{equation}
where the normalization constants depend on the energies
$C_{n_l,\pm}:=[(E_{n_l}\pm mc^2)/2E_{n_l}]^{1/2}$. The
transformation in Eq.~\eqref{unitary_transformation_left} can be
rewritten in the monochromatic scenario as
\begin{equation}
\label{unitary_single_freq_left}
 \text{U}_{\alpha}=\ee^{-\frac{\alpha\tilde{\mu}}{2}\left(-a_ra_l+a_r^{\dagger}a^{\dagger}_{l}\right)},
\end{equation}
where
$\tilde{\mu}:=(\Delta/\tilde{\Delta}+\tilde{\Delta}/\Delta)/2$.
This transformation can be immediately related to a two-mode
squeezing operator in the context of Quantum Optics with squeezing
parameter
$z:=-\alpha\tilde{\mu}/2\in\mathbb{R}$~\cite{two_mode_squeezing}.
The action of such an squeezing
operator~\eqref{unitary_single_freq_left} over left-handed chiral
Fock states gives rise to SU(1,1) coherent states
$\ket{z,n_l}:=\text{U}_{\alpha}^{\dagger}\ket{n_l}\ket{\text{vac}}_r$~\cite{perelomov}(see
Appendix~\ref{apendix_a} for some details)
\begin{equation}
\label{su11_left_coherent_states}
\ket{z,n_l}=\mathcal{N}_{n_l}\sum_{m=0}^{\infty}\sqrt{\frac{(m+n_l)!}{n_l!m!}}(-1)^m\text{tanh}^m|z|
\ket{m+n_l,m},
\end{equation}
where we have introduced the normalization constant
$\mathcal{N}_{n_l}:=\text{cosh}^{-(n_l+1)}|z|$. Since right-handed
operators are not present in the effective Hamiltonian of
Eq.~\eqref{transformed_AJC_left}, and consequently do not
participate in the relativistic dynamics, we have chosen the
right-handed vacuum for simplicity. The state corresponding to
$n_l=0$ can be
 identified with the two-mode squeezed vacuum state, an archetypical state in the field of Quantum Optics that becomes
 highly non-classical for a large squeezing parameter. Therefore,
 the energy eigenstates of the relativistic fermion in
 Eq.~\eqref{left_squeezed_eigenstates} can be expressed as
 \begin{equation}
\label{left_eigenstates_coh_doublet}
 \ket{\pm
E_{n_l}}=C_{n_l,\pm}\ket{z,n_l+1}\ket{\chi_{\uparrow}}\mp\ii
C_{n_l,\mp}\ket{z,n_l}\ket{\chi_{\downarrow}},
\end{equation}
which is identical to an AJC-like doublet where SU(1,1) coherent
states are entangled with the relativistic spinors. Such an state
possesses remarkable non-classical properties, such as spin-orbit
entanglement or sub-Poissonian statistics. With respect to the
weak magnetic field eigenstates in
Eqs.~\eqref{left_handed_eigenstates}, these states present certain
novel features, such as an inter-mode chiral entanglement, or
bosonic statistics which depend on the coupling strength
$\tilde{\xi}/\xi$. All these interesting properties will be
described in detail in forthcoming sections. We shall be also
interested in the properties of the fermionic ground state
$E_{g}=mc^2$
 \begin{equation}
\label{left_ground_state}
\ket{g}=\ket{z,0}\ket{\chi_{\uparrow}}=\frac{1}{\text{cosh}|z|}\sum_{m=0}^{\infty}(-1)^m\text{tanh}^m|z|\ket{m,m}\ket{\chi_{\uparrow}},
\end{equation}
which can be notably interpreted as a spin-up squeezed vacuum
state, where the squeezing parameter $z=-\alpha\tilde{\mu}/2$
depends on the relative coupling strengths $\xi,\tilde{\xi}$.

\subsection{Right-handed regime $\tilde{\xi}>\xi$}

In this case, the original Hamiltonian in
Eqs.~\eqref{JC-AJC-hamiltonian}-\eqref{AJC_left} is transformed
into a Jaynes-Cummings Hamiltonian under the action of
\begin{equation}
\label{unitary_transformation_right} \text{U}_{\tilde{\alpha}}:=
  \ee^{\tilde{\alpha}\left(a_l\tilde{a}_r-\tilde{a}^{\dagger}_ra_l^{\dagger}\right)},
\end{equation}
where the parameter $\tilde{\alpha}$  in this regime becomes
\begin{equation}
\tilde{\alpha}:=\frac{1}{\lambda}\text{arctanh}\left(\frac{\lambda\sqrt{\omega}}{\sqrt{\tilde{\omega}}+\mu\sqrt{\omega}}\right).
\end{equation}
The chiral operators are once again transformed according to
Eqs.~\eqref{operators_transformation} when the substitution
$\alpha\to\tilde{\alpha}$ is performed. In this situation, the
transformed Hamiltonian becomes a single-mode Jaynes-Cummings term
\begin{equation}
\label{transformed_JC_right}
H_{\text{JC}}^{\circlearrowleft}(g'_r):=\text{U}_{\tilde{\alpha}}H\text{U}^{\dagger}_{\tilde{\alpha}}=\delta\sigma_z+g_r'\sigma^{+}a_r+(g_r')^{*}\sigma^{-}a^{\dagger}_r,
\end{equation}
where the new coupling strength is  $g'_r:=\ii
mc^2\sqrt{2\zeta_r}$, and
$\zeta_r(\xi,\tilde{\xi}):=\tilde{\xi}-\xi+2\mu(\tilde{\xi}\xi)^{1/2}$.
Analogously to the left-handed regime $\tilde{\xi}<\xi$, we obtain
a single-mode Hamiltonian which can be easily diagonalized
following the same procedure as in the strong magnetic field
limit, and provides the solution to the complete bichromatic
interaction in Eq.~\eqref{JC-AJC-hamiltonian}. In this sense, the
energy spectrum becomes
\begin{equation}
\label{transformed_right_handed_energy} \tilde{E}=\pm
\tilde{E}_{\tilde{n}_r}=\pm
mc^2\sqrt{1+2\zeta_r(\xi,\tilde{\xi})(\tilde{n}_r+1)},
\end{equation}
where $\tilde{n}_r=0,1...$ represents the number of right-handed
quanta. Applying the unitary transform in
Eq.~\eqref{unitary_transformation_right} to the single-mode
eigenstates, one obtains the corresponding eigenstates of the
complete original Hamiltonian
\begin{equation}
\label{right_squeezed_eigenstates} \ket{\pm
\tilde{E}_{\tilde{n}_r}}=\text{U}_{\tilde{\alpha}}^{\dagger}\left(\tilde{C}_{\tilde{n}_r,\pm}\ket{\tilde{n}_r}\ket{\chi_{\uparrow}}\mp\ii
\tilde{C}_{\tilde{n}_r,\mp}\ket{\tilde{n}_r+1}\ket{\chi_{\downarrow}}\right),
\end{equation}
where $\tilde{C}_{\tilde{n}_r,\pm}:=[(\tilde{E}_{\tilde{n}_r}\pm
mc^2)/2\tilde{E}_{\tilde{n}_r}]^{1/2}$ . In this regime, the
transformation in Eq.~\eqref{unitary_transformation_right} in the
single-frequency domain becomes
\begin{equation}
\label{unitary_single_freq_right}
 \text{U}_{\tilde{\alpha}}=\ee^{-\frac{\tilde{\alpha}\tilde{\mu}}{2}\left(-\tilde{a}_r\tilde{a}_l+\tilde{a}_r^{\dagger}\tilde{a}^{\dagger}_{l}\right)},
\end{equation}
which can be reinterpreted once more as a two-mode squeezing
operator with a different squeezing parameter
$\tilde{z}:=-\tilde{\alpha}\tilde{\mu}/2\in\mathbb{R}$. This
operator transforms the right-handed chiral Fock states
$\ket{\tilde{n}_r}$ into SU(1,1) coherent states
$\ket{\tilde{z},\tilde{n}_r}:=\text{U}_{\tilde{\alpha}}^{\dagger}\ket{\text{vac}}_l\ket{\tilde{n}_r}$,
where we have chosen the left-handed vacuum for simplicity (see
Appendix~\ref{apendix_a})
\begin{equation}
\label{right_coh_states}
\ket{\tilde{z},\tilde{n}_r}=\mathcal{N}_{\tilde{n}_r}\sum_{\tilde{m}=0}^{\infty}\sqrt{\frac{(\tilde{m}+\tilde{n}_r)!}{\tilde{n}_r!\tilde{m}!}}(-1)^{\tilde{m}}\text{tanh}^{\tilde{m}}|\tilde{z}|
\ket{\tilde{m}+\tilde{n}_r,\tilde{m}},
\end{equation}
where we have introduced the following normalization constant
$\mathcal{N}_{\tilde{n}_r}:=\text{cosh}^{-(\tilde{n}_r+1)}|\tilde{z}|$.
Such coherent states in Eq.~\eqref{right_coh_states} appear in
doublets in the fermionic eigenstates of
Eq.~\eqref{left_ground_state}
 \begin{equation}
\label{right_eigenstates_coh_doublet}
 \ket{\pm
\tilde{E}_{\tilde{n}_r}}=\tilde{C}_{\tilde{n}_r,\pm}\ket{\tilde{z},\tilde{n}_r}\ket{\chi_{\uparrow}}\mp\ii
\tilde{C}_{\tilde{n}_r,\mp}\ket{\tilde{z},\tilde{n}_r+1}\ket{\chi_{\downarrow}}.
\end{equation}
Once again, one can identify these eigenstates as JC-like doublets
where SU(1,1) coherent states become entangled with the
relativistic spin degree of freedom. Conversely to the fermionic
ground state of the left-handed regime in
Eq.~\eqref{left_ground_state}, the ground state in this regime
$E_{\tilde{g}}=mc^2\sqrt{1+2\zeta_r}, $ cannot be expressed solely
by the two-mode squeezed vacuum, but rather by
\begin{equation}
\label{right_ground_state}
\ket{\tilde{g}}=\tilde{C}_{0+}\ket{\tilde{z},0}\ket{\chi_{\uparrow}}-\ii
\tilde{C}_{0-}\ket{\tilde{z},1}\ket{\chi_{\downarrow}},
\end{equation}
which besides the two-mode squeezed vacuum, also includes the
$\tilde{n}_r=1$ SU(1,1) coherent state.

\subsection{Critical regime $\tilde{\xi}=\xi$}

In this regime, the effective Hamiltonian can be directly obtained
from Eq.~\eqref{dirac_oscillator_magnetic_field_2D}
\begin{equation}
H^{\text{c}}:=H_{\text{2D}}^{\text{free}}=c\sum_{j=1}^2\sigma_jp_j+\sigma_z
mc^2,
\end{equation}
which corresponds to the Dirac Hamiltonian of a two-dimensional
free fermion. In the critical regime $\tilde{\xi}=\xi$, the
magnetic field coupling cancels the effect of the Dirac string
coupling, and the fermion behaves as a free relativistic
particles. The critical energy spectrum becomes
\begin{equation}
E^{\text{c}}=\pm
E^{\text{c}}_{\textbf{p}}=\pm\sqrt{m^2c^4+\textbf{p}^2c^2},
\end{equation}
where $\textbf{p}=(p_x,p_y)$ stands for the two-dimensional
fermion momentum. The corresponding eigenstates can be described
as follows
\begin{equation}
\label{critical_eigenstates} \ket{\pm
E^{\text{c}}_{\textbf{p}}}=\sqrt{\frac {mc^2\pm
E^{\text{c}}_{\textbf{p}}}{\pm
2E^{\text{c}}_{\textbf{p}}}}\left(\ket{\chi_{\uparrow}}+\frac{c(p_x+\ii
p_y)}{mc^2\pm
E^{\text{c}}_{\textbf{p}}}\ket{\chi_{\downarrow}}\right)\ket{\textbf{p}},
\end{equation}
where $\ket{\textbf{p}}:=\ket{p_x,p_y}$ are two-dimensional plane
wave solutions. Note that these solutions describe the
relativistic fermion at the critical point $\tilde{\xi}=\xi$, but
any small perturbation of the magnetic field or the Dirac string
coupling will dramatically change the system properties. As we
discuss below, this dramatic change shares many analogies with a
quantum phase transition.

 We have thus provided a complete solution of the
relativistic Hamiltonian that describes the properties of a Dirac
oscillator subjected to an additional constant magnetic field. A
unitary transformation that connects the bichromatic full
Hamiltonian with single-mode JC-like interactions has been
described in
Eqs.~\eqref{unitary_transformation_left},~\eqref{unitary_transformation_right}.
With the aid of such transformation, one can obtain the exact
energy spectrum for all possible values of the coupling parameters
$\xi,\tilde{\xi}$, which responds to the analytical expressions in
Eqs.~\eqref{transformed_left_handed_energy},\eqref{transformed_right_handed_energy}.
In this regard, one can study the transition between the weak
$\tilde{\xi}/\xi\to 0$ and strong $\tilde{\xi}/\xi\to \infty$
coupling regimes which endows the relativistic system with a
chiral symmetry change, as discussed in the previous section.

From a practical point of view, the magnetic field dependence of
this relativistic system might be understood as an accessible
method to prepare a certain relativistic state with specific
properties. An adiabatic control of the applied magnetic field
opens up the possibility of controlling a wide range of properties
of this relativistic fermion, such as its chirality, squeezing,
phonon statistics, and entanglement, which shall be discussed
below.

\section{Chirality Quantum Phase Transition}
\label{sectionV}

 In previous sections, we have described how the value of the
 relative coupling strength $\tilde{\xi}/\xi$ is responsible for
 the chirality of the system, which can be right- or left-handed
 by an appropriate tuning of the coupling strengths.  This drastic
 modification can only be driven by means of quantum fluctuations,
 and therefore a quantum phase transitions occurs in this
 relativistic system. In this section we study several properties
which are clear signatures of a quantum phase transition
phenomenon, such as the vanishing of the gap at the critical
point, the existence of an order parameter which takes on
different values in the distinct chiral phases, the divergence of
quantum fluctuations and the maximization of entanglement across
the critical point.

\subsection{Energy level crossing}

\begin{figure}[!hbp]

\centering

\begin{overpic}[width=8.5cm]{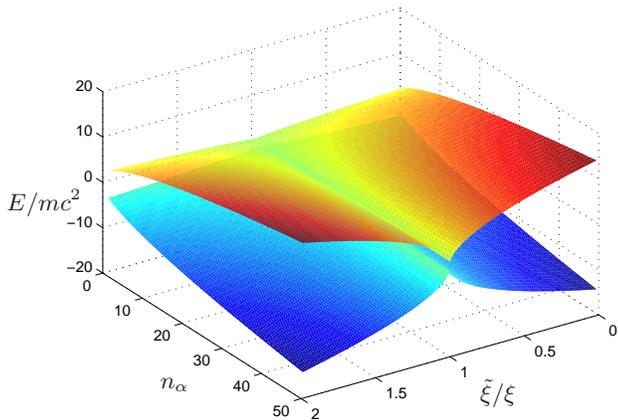}
\put(-2,38){{$E/mc^2$}}
\put(22,10){$n_{\alpha}$}\put(72,8){$\tilde{\xi}/\xi$}

\end{overpic}
\caption{Fermionic energy spectrum as a function of the relative
coupling strengths $\tilde{\xi}/\xi$ and the number of chiral
phonons $n_{\alpha}$ ($n_{\alpha}=n_l$ if $\tilde{\xi}<\xi$, and
$n_{\alpha}=\tilde{n}_r$ if $\tilde{\xi}>\xi$). The two energy
sheets correspond to positive- and negative-energy solutions
$\ket{\pm E_{n_l}}$ for $\tilde{\xi}<\xi$, and $\ket{\pm
\tilde{E}_{\tilde{n}_r}}$ for $\tilde{\xi}>\xi$.
}\label{energias_2D}

\end{figure}

Here, we study the properties of the relativistic energy spectrum
described in
 Eqs.~\eqref{transformed_left_handed_energy},\eqref{transformed_right_handed_energy},
 which are represented in fig.~\ref{energias_2D}. Here we represent the energies of different eigenstates  $\ket{\pm E_{n_l}},\ket{\pm \tilde{E}_{\tilde{n}_r}}$
with respect to the relative strength $\tilde{\xi}/\xi$.

\begin{figure}[!hbp]

\centering

\begin{overpic}[width=8.5cm]{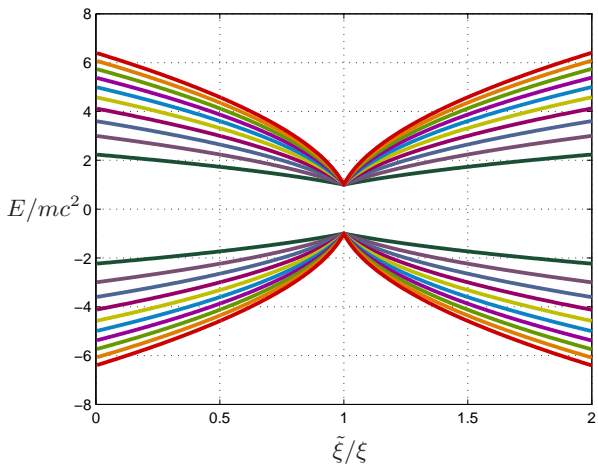}
\put(-1,38){{$E/mc^2$}} \put(50,0){$\tilde{\xi}/\xi$}
\end{overpic}

\caption{Energy levels for the first excited eigenstates as a
function of the relative coupling strengths $\tilde{\xi}/\xi$,
showing the non-analytic behavior at $(\tilde{\xi}/\xi)_c=1$.
}\label{energias}

\end{figure}

We clearly observe two crucial signatures of a quantum phase
transition (see also fig.~\ref{energias} for a two-dimensional
sector of the above energy spectrum):

\begin{itemize}
\item The energy spectrum is non-analytical for a critical coupling
$(\tilde{\xi}/\xi)_c=1$,

\item The system becomes gapless at this
critical point $\Delta E \to 0$ when
$|(\tilde{\xi}/\xi)-(\tilde{\xi}/\xi)_c|\to 0$.

\end{itemize}

Furthermore, we can obtain an analytical expression of the gap
close to the critical point. In terms of the coupling strengths
 $g_r$ and $g_l$
in Eqs.~\eqref{JC_right}-\eqref{AJC_left}, we obtain the following
universal scaling law for the energy gap close to the critical
point $(g_r/g_l)_c=(\tilde{\xi}/\xi)_c^{1/2}=1$
\begin{equation}
\Delta E\sim
mc^2\left|\frac{g_r}{g_l}-\left(\frac{g_r}{g_l}\right)_c\right|=:mc^2\left|\frac{g_r}{g_l}-\left(\frac{g_r}{g_l}\right)_c\right|^{z\nu},
\end{equation}
and we can readily identify the scaling exponents $z\nu=1$. The
critical exponent $z$ is a dynamical exponent related to the decay
of the system fluctuations with time, and in a Lorentz invariant
model is expected to be $z=1$. Consequently, the remaining
exponent, which characterizes the scaling of the correlation
length close to the phase transition, becomes $\nu=1$ in this
relativistic model.

\begin{figure}[!hbp]

\centering

\begin{overpic}[width=8.5cm]{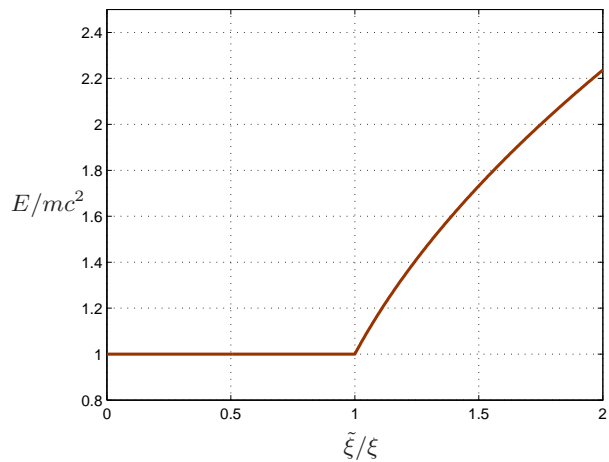}
\put(-2,38){{$E/mc^2$}} \put(50,0){$\tilde{\xi}/\xi$}
\end{overpic}

\caption{Ground state energy as a function of the relative
coupling $\tilde{\xi}/\xi$. As the magnetic filed is increased,
the ground state energy changes from $E_g=mc^2\to
E_{\tilde{g}}=mc^2\sqrt{1+2\zeta_r}$, which becomes non-analytical
at the critical point.}\label{gs_energias}

\end{figure}

Finally, it is also important in quantum phase transitions to
study the properties of the system ground state. In this case, the
fermionic ground state in the left-handed regime is described by
Eq.~\eqref{left_ground_state}, whilst the ground state in the
right-handed scenario becomes that of
Eq.~\eqref{right_ground_state}. The energy of such ground state is
non-analytical at the critical point a follows from
fig.~\ref{gs_energias}.

\subsection{Order parameter}

Another clear signature of quantum phase transitions is the
appearance of an order parameter, a physical magnitude which
acquires different values in the phases involved and becomes
indeterminate at the critical point. Therefore, an order parameter
is able to witness the abrupt change in the properties of the
system. In this relativistic scenario, the order parameter turns
out to be the $z$-component of the orbital angular momentum
$L_z=xp_y-yp_x$, an observable which is intimately related to the
symmetry properties of the system, since $J_z=L_z+S_z$ is a
conserved quantity. Note that this operator is also connected to
the system chirality, since left-handed states fulfill $\langle
L_z \rangle_l\leq0$ whereas right-handed states fulfill $\langle
L_z \rangle_r\geq0$. In order to evaluate the angular momentum
expectation value, it is useful to express the unitary
transformations in
Eqs.~\eqref{unitary_single_freq_left},\eqref{unitary_single_freq_right}
in terms of canonical conjugate position and momentum operators
\begin{equation}
\label{unitary_transformation_position_momentum}
\begin{split}
U_{\alpha}&=\ee^{\ii\frac{\alpha\tilde{\mu}}{4\hbar}\left(xp_x+p_xx+yp_y+p_yy\right)},\hspace{2ex}
\text{if} \hspace{2ex} \tilde{\xi}<\xi,\\
U_{\tilde{\alpha}}&=\ee^{\ii\frac{\tilde{\alpha}\tilde{\mu}}{4\hbar}\left(xp_x+p_xx+yp_y+p_yy\right)},\hspace{2ex}
\text{if} \hspace{2ex} \tilde{\xi}>\xi.
\end{split}
\end{equation}
Using the transformations in
Eqs.~\eqref{unitary_transformation_position_momentum}, one can
obtain the expressions of the transformed position and momentum
operators
\begin{equation}
\label{position_momentum_operators}
\begin{split}
&\text{U}_{\alpha}x_i\text{U}^{\dagger}_{\alpha}=\ee^{\frac{\alpha\tilde{\mu}}{2}}x_i,
\hspace{3ex}\text{U}_{\alpha}p_i\text{U}^{\dagger}_{\alpha}=\ee^{-\frac{\alpha\tilde{\mu}}{2}}p_i,
\hspace{2ex}
\text{if} \hspace{2ex} \tilde{\xi}<\xi,\\
&\text{U}_{\tilde{\alpha}}x_i\text{U}^{\dagger}_{\tilde{\alpha}}=\ee^{\frac{\tilde{\alpha}\tilde{\mu}}{2}}x_i,
\hspace{3ex}\text{U}_{\tilde{\alpha}}p_i\text{U}^{\dagger}_{\tilde{\alpha}}=\ee^{-\frac{\tilde{\alpha}\tilde{\mu}}{2}}p_i,
\hspace{2ex} \text{if} \hspace{2ex} \tilde{\xi}>\xi,
\end{split}
\end{equation}
with $i=x,y$, which leads to the corresponding relations for the
orbital angular momentum
\begin{equation}
\label{ang_momentum_transformation}
\begin{split}
\text{U}_{\alpha}L_z\text{U}^{\dagger}_{\alpha}=\text{U}_{\alpha}(xp_y-yp_x)\text{U}^{\dagger}_{\alpha}=L_z,\hspace{2ex}
\text{if} \hspace{2ex} \tilde{\xi}<\xi,\\
\text{U}_{\tilde{\alpha}}L_z\text{U}^{\dagger}_{\tilde{\alpha}}=\text{U}_{\tilde{\alpha}}(xp_y-yp_x)\text{U}^{\dagger}_{\tilde{\alpha}}=L_z,\hspace{2ex}
\text{if} \hspace{2ex} \tilde{\xi}>\xi.\\
\end{split}
\end{equation}

\begin{figure}[!hbp]

\centering

\begin{overpic}[width=8.5cm]{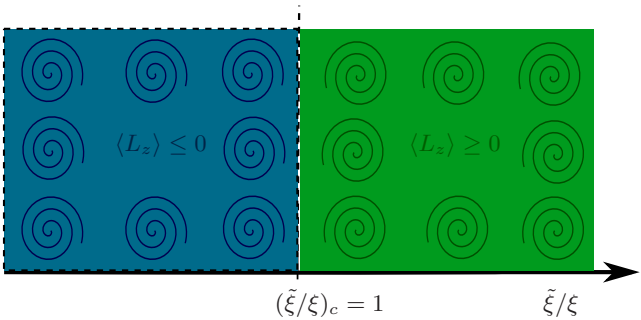}
\put(18,30){{\textcolor[rgb]{0.00,0.14,0.28}{$\langle
L_z\rangle\leq0$}}}
\put(64,30){\textcolor[rgb]{0.00,0.37,0.00}{$\langle
L_z\rangle\geq0$}}\put(43,5){$(\tilde{\xi}/\xi)_c=1$}
\put(85,5){$\tilde{\xi}/\xi$}
\end{overpic}

\caption{Mean value of the $z$-component of the orbital angular
momentum across the critical point. Note that, as the critical
region is traversed, the sign of this order parameter changes and
indicates that the phase transition has
occurred.}\label{ang_momentum_order_parameter}

\end{figure}
We observe from the expressions in
Eqs.~\eqref{ang_momentum_transformation} that the $z$-component of
the orbital angular momentum operator is not altered by the
squeezing transformation. In the language of Lie algebras, this
property is equivalent to $L_z$ being the Casimir operator
associated to the underlying SU(1,1) algebra (see
Apendix~\ref{apendix_a}). Therefore, we can easily obtain its
expectation value in the eigenstates of
Eqs.~\eqref{left_squeezed_eigenstates},\eqref{right_squeezed_eigenstates}
\begin{equation}
\label{order_parameter}
\begin{split}
\langle L_z\rangle_{\ket{\pm
E_{n_l}}}&=-\hbar\left(n_l+\frac{E_{n_l}\pm
mc^2}{2E_{n_l}}\right)\leq0,\hspace{2ex} \text{if} \hspace{2ex}
\tilde{\xi}<\xi,\\
\langle L_z\rangle_{\ket{\pm
\tilde{E}_{\tilde{n}_r}}}&=+\hbar\left(\tilde{n}_r+\frac{\tilde{E}_{\tilde{n}_r}\mp
mc^2}{2\tilde{E}_{\tilde{n}_r}}\right)\geq0,\hspace{2ex} \text{if}
\hspace{2ex} \tilde{\xi}>\xi,
\end{split}
\end{equation}
which show how the orbital angular momentum takes on negative
values in the left-handed chiral phase, whereas it attains
positive values in the right-handed chiral regime (see
fig.~\ref{ang_momentum_order_parameter}). Consequently, the
orbital angular momentum plays the role of an order parameter
which witnesses the quantum phase transition and macroscopically
reveals such effect by a change of its sign.

\subsection{Divergence of quantum fluctuations  }

In the vicinity of a critical point, the gap of the system becomes
negligible and excitations can be easily produced (see
fig.~\ref{energias}). Under such conditions, the system becomes
highly fluctuating. As discussed previously, a quantum phase
transition can only be driven by quantum fluctuations which lead
to an abrupt change in the physical properties of the system when
the critical region is crossed. In our case, it is possible to
calculate analytically the divergences in the quantum fluctuations
of the fermion position $\Delta x_i=\sqrt{\langle
x_i^2\rangle-\langle x_i\rangle^2}$.

\begin{figure}

\centering

\begin{overpic}[width=8.5cm]{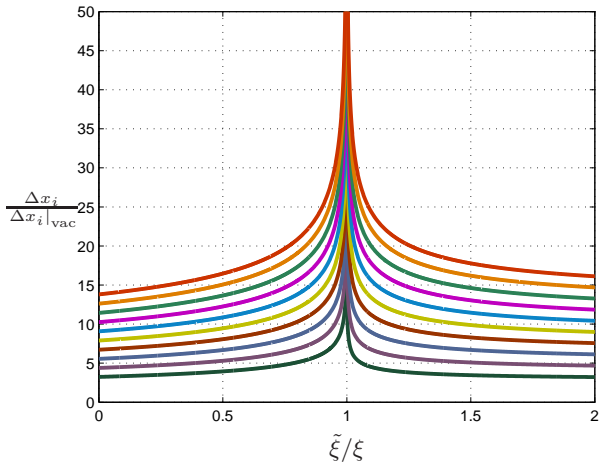}
\put(-2,38){{$\frac{\Delta x_i}{\left.\Delta
x_i\right|_{\text{vac}}}$}}\put(49,0){$\tilde{\xi}/\xi$}

\end{overpic}
\caption{Quantum fluctuations in the fermion position $\Delta x_i$
for the first excited positive-energy eigenstates $\ket{+
E_{n_l}},\ket{+ \tilde{E}_{\tilde{n}_r}}$ as a function of the
relative coupling strengths $\tilde{\xi}/\xi$. We note that the
negative-energy eigenstates display similar fluctuations in
position.}\label{position_fluctuations_2D}

\end{figure}

We shall not only focus in the ground state fluctuations in
Eqs.~\eqref{left_ground_state},\eqref{right_ground_state}, but we
shall investigate the quantum fluctuations of the whole energy
eigenstates in
Eqs.~\eqref{left_eigenstates_coh_doublet},\eqref{right_eigenstates_coh_doublet}.
Using the transformations described in
Eq.~\eqref{position_momentum_operators}, we obtain
\begin{equation}
\begin{split}
&\left.\Delta x_i\right|_{\ket{\pm E_{n_l}}} =\left.\Delta
x_i\right|_{\text{vac}}\sqrt{\eta_{\pm}}\ee^{+\frac{\alpha\tilde{\mu}}{2}},
\hspace{2ex} \text{if} \hspace{2ex} \tilde{\xi}<\xi,\\
&\left.\Delta x_i\right|_{\ket{\pm \tilde{E}_{\tilde{n}_r}}}
=\left.\tilde{\Delta}
x_i\right|_{\text{vac}}\sqrt{\tilde{\eta}_{\mp}}\ee^{+\frac{\tilde{\alpha}\tilde{\mu}}{2}},
\hspace{2ex} \text{if} \hspace{2ex} \tilde{\xi}>\xi,
\end{split}
\end{equation}
where $i=x,y$, $\Delta x_i|_{\text{vac}}:=\Delta/\sqrt{2}$ stand
for the fluctuations of the vacuum in modes $a_r,a_l$, whereas
$\tilde{\Delta} x_i|_{\text{vac}}:=\tilde{\Delta}/\sqrt{2}$ are
the vacuum fluctuations in modes  $\tilde{a}_r,\tilde{a}_l$. We
have also introduced $\eta_{\pm}:=n_l+3/2\pm mc^2/2E_{n_l}$, and
$\tilde{\eta}_{\pm}:=\tilde{n}_r+3/2\pm
mc^2/2\tilde{E}_{\tilde{n}_r}$, which depend on the number of
quanta. We have represented in fig.~\ref{position_fluctuations_2D}
the fluctuations  in the fermion position for different
eigenstates as the relative coupling $\tilde{\xi}/\xi$ is varied.
We observe how these fluctuations diverge at the critical point
$(\tilde{\xi}/\xi)_c=1$ for all the eigenstates, representing yet
another signature of a quantum phase transition.

\begin{figure}

\centering

\begin{overpic}[ width=8.5cm]{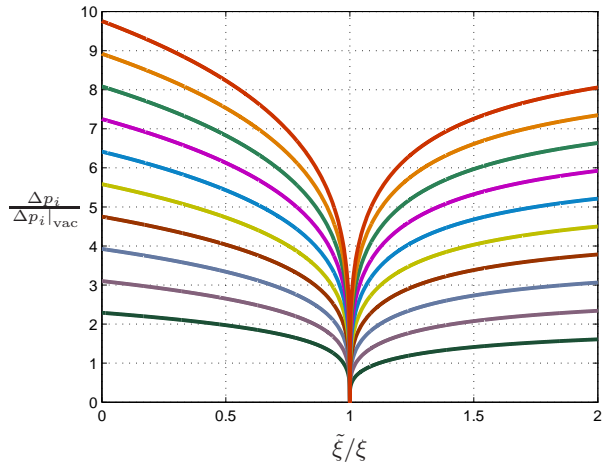}
\put(-2,38){{$\frac{\Delta p_i}{\left.\Delta
p_i\right|_{\text{vac}}}$}}\put(49,0){$\tilde{\xi}/\xi$}

\end{overpic}
\caption{Quantum fluctuations in the fermion momentum$\Delta p_i$
for the first excited positive-energy eigenstates $\ket{+
E_{n_l}},\ket{+\tilde{E}_{\tilde{n}_r}}$ as a function of the
relative coupling strengths $\tilde{\xi}/\xi$. We note that the
negative-energy eigenstates display similar fluctuations in
momentum.}\label{momentum_fluctuations_2D}

\end{figure}

Nonetheless, we have remarked in
Eqs.~\eqref{unitary_single_freq_left},\eqref{unitary_single_freq_right}
that the unitary transformation involved corresponds to a
squeezing transformation in the language of quantum optics. As a
consequence, a squeezing of certain fluctuations must also become
apparent for a certain observable. This is the case of the
fermionic momentum, whose fluctuations $\Delta p_i=\sqrt{\langle
p_i^2\rangle-\langle p_i\rangle^2}$ for different eigenstates are
the following

\begin{equation}
\begin{split}
&\left.\Delta p_i\right|_{\ket{\pm E_{n_l}}} =\left.\Delta
p_i\right|_{\text{vac}}\sqrt{\eta_{n_l,\pm}}\ee^{-\frac{\alpha\tilde{\mu}}{2}},
\hspace{2ex} \text{if} \hspace{2ex} \tilde{\xi}<\xi,\\
&\left.\Delta p_i\right|_{\ket{\pm \tilde{E}_{\tilde{n}_r}}}
=\left.\tilde{\Delta}
p_i\right|_{\text{vac}}\sqrt{\eta_{\tilde{n}_r,\mp}}\ee^{-\frac{\tilde{\alpha}\tilde{\mu}}{2}},
\hspace{2ex} \text{if} \hspace{2ex} \tilde{\xi}>\xi,
\end{split}
\end{equation}
where $\Delta p_i|_{\text{vac}}=\hbar/\Delta\sqrt{2}$, and
$\tilde{\Delta} p_i|_{\text{vac}}=\hbar/\tilde{\Delta}\sqrt{2}$
stand for the vacuum fluctuations in each regime. Remarkably, the
fluctuations in the momentum vanish as the system approaches the
critical point (see fig.~\ref{momentum_fluctuations_2D}).
Therefore, the squeezing occurs in the relativistic momentum.

With these observations, we have a full description of the system
for any value of the relative coupling:

\begin{itemize}
\item \textbf{ Left-handed regime} $\tilde{\xi}<\xi$.-- The relativistic
eigenstates are described by an AJC-like doublet of SU(1,1)
coherent states entangled with Pauli
spinors~\eqref{left_eigenstates_coh_doublet}, which are  rotating
clockwise (see Eq.~\eqref{order_parameter}).
\item \textbf{ Critical regime} $\tilde{\xi}=\xi$.-- The
critical properties are described by those of a relativistic free
fermion.
\item \textbf{ Right-handed regime} $\tilde{\xi}>\xi$.-- The relativistic
eigenstates are described by an JC-like doublet of SU(1,1)
coherent states entangled with Pauli
spinors~\eqref{right_eigenstates_coh_doublet}, which are  rotating
counterclockwise (see Eq.~\eqref{order_parameter}).
\end{itemize}

\subsection{Phonon statistical properties}

In this section we study the statistical properties of the chiral
phonon distribution. The statistical properties of a bosonic
ensemble can be classified according to its quantum fluctuations:
\begin{itemize}
\item \textbf{Poissonian statistics}.-- In this case, the phonon
number distribution is a Poissonian random variable. Quantum
fluctuations in the number of bosons fulfill $\Delta
n=\sqrt{\langle n\rangle}$, where $n=a^{\dagger}a$ is the bosonic
number operator, and $\Delta n=\sqrt{\langle n^2\rangle-\langle
n\rangle^2}$.

\item \textbf{Super-Poissonian statistics}.-- In this regime, the
quantum fluctuations of the bosonic ensemble are bigger with
respect to the Poissonian distribution $\Delta n>\sqrt{\langle
n\rangle}$, and therefore the phonons are said to be noisier.

\item \textbf{Sub-Poissonian statistics}.-- The quantum nature of a bosonic
ensemble allows a further possibility, which is an evidence of
non-classical behaviour, that of sub-Poissonian statistics. Under
these circumstances, the statistical fluctuations become lower
than those of a Poissonian distribution $\Delta n<\sqrt{\langle
n\rangle}$, and consequently the bosons are said to be quieter
quantum entities.
\end{itemize}

\begin{figure}

\centering

\begin{overpic}[width=8.5cm]{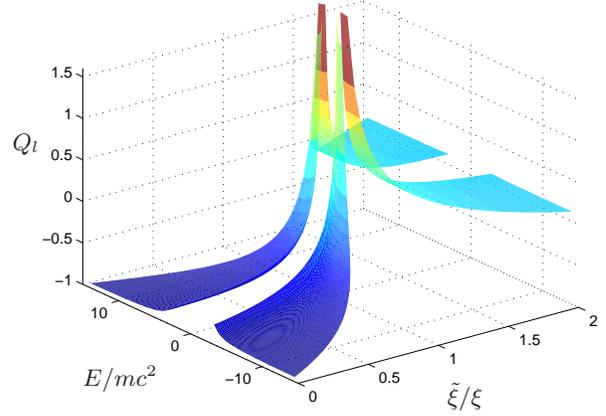}
\put(2,45){{$Q_l$}}
\put(13,8){$E/mc^2$}\put(70,5){$\tilde{\xi}/\xi$}

\end{overpic}
\caption{Left-handed Mandel parameter $Q_l$ for the energy
eigenstates $\{\ket{\pm E_{n_l}},\ket{\pm
\tilde{E}_{\tilde{n}_r}}\}$ as a function of the relative coupling
strengths $\tilde{\xi}/\xi$. Note that as the magnetic field is
raised, the ensemble of left-handed phonons changes from sub- to
super-Poissonian statistics.}\label{left_statistics_2D}

\end{figure}

\begin{figure}

\centering

\begin{overpic}[ width=8.5cm]{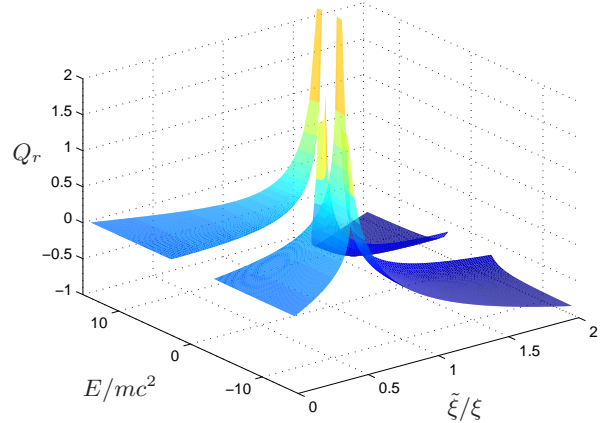}
\put(2,45){{$Q_r$}}
\put(13,8){$E/mc^2$}\put(70,5){$\tilde{\xi}/\xi$}

\end{overpic}
\caption{Right-handed Mandel parameter $Q_r$ for the energy
eigenstates $\{\ket{\pm E_{n_l}},\ket{\pm
\tilde{E}_{\tilde{n}_r}}\}$ as a function of the relative coupling
strengths $\tilde{\xi}/\xi$. Note that as the magnetic field is
raised, the ensemble of right-handed phonons changes from super-
to sub-Poissonian statistics.}\label{right_statistics_2D}

\end{figure}

In this section we show how the relativistic phonon distribution
attains different statistical properties depending on the relative
coupling parameter $\tilde{\xi}/\xi$. In particular,we show how
the quantum noise of the chiral phonon distribution changes from
sub-Poissonian to super-Poissonian as the critical point is
crossed. In order to quantify the Poissonian character of the
chiral boson ensemble, we shall make use of the chiral Mandel $Q$
parameters, defined as follows
\begin{equation}
\label{mandel_param} Q_{r}=\frac{(\Delta n_r)^2}{\langle
n_r\rangle}-1,\hspace{2ex} Q_{l}=\frac{(\Delta n_l)^2}{\langle
n_l\rangle}-1.
\end{equation}
A positive sign in such parameters reveals the super-Poissonian
nature of the bosonic ensemble, whereas a negative sign shows the
sub-Poissonian bosonic statistics which stress the non-classical
nature of the ensemble. The corresponding fluctuations in the
system
eigenstates~\eqref{left_eigenstates_coh_doublet},\eqref{right_eigenstates_coh_doublet},
can be calculated using the expressions in
Eqs.~\eqref{left_coh_exp_values},\eqref{right_coh_exp_values} in
Appendix~\ref{apendix_a}. In the left-handed regime
$\tilde{\xi}<\xi$, we have found
\begin{equation}
\label{left_phonon_statistics}
\begin{split}
&\langle n_r\rangle_{|\pm E_{n_l}\rangle}
=\eta_{\pm}\text{sinh}^2|z|,\\
&\langle n_l\rangle_{|\pm E_{n_l}\rangle}
=\eta_{\pm}\text{cosh}^2|z|-1,\\
&\left.\Delta n_r\right|^2_{|\pm E_{n_l}\rangle}
=\eta_{\pm}\text{sinh}^2|z|\text{cosh}^2|z|+\kappa\text{sinh}^4|z|,\\
&\left.\Delta n_l\right|^2_{|\pm E_{n_l}\rangle}
=\eta_{\pm}\text{sinh}^2|z|\text{cosh}^2|z|+\kappa\text{cosh}^4|z|,\\
\end{split}
\end{equation}
where the parameters $\eta_{\pm}$ have been previously introduced,
and $\kappa:=\frac{1}{4}(1-m^2c^4/{E^{'2}_{n_l}})$. Analogously,
in the right-handed case $\tilde{\xi}>\xi$
\begin{equation}
\label{right_phonon_statistics}
\begin{split}
&\langle \tilde{n}_r\rangle_{|\pm \tilde{E}_{\tilde{n}_r}\rangle}=\tilde{\eta}_{\mp}\text{cosh}^2|\tilde{z}|-1,\\
&\langle \tilde{n}_l\rangle_{|\pm \tilde{E}_{\tilde{n}_r}\rangle}=\tilde{\eta}_{\mp}\text{sinh}^2|\tilde{z}|,\\
&\left.\Delta\tilde{n}_r\right|^2_{|\pm \tilde{E}_{\tilde{n}_r}\rangle}=\tilde{\eta}_{\mp}\text{sinh}^2|\tilde{z}|\text{cosh}^2|\tilde{z}|+\tilde{\kappa}\text{cosh}^4|\tilde{z}|,\\
&\left.\Delta\tilde{n}_l\right|^2_{|\pm \tilde{E}_{\tilde{n}_r}\rangle}=\tilde{\eta}_{\mp}\text{sinh}^2|\tilde{z}|\text{cosh}^2|\tilde{z}|+\tilde{\kappa}\text{sinh}^4|\tilde{z}|,\\
\end{split}
\end{equation}
where $\tilde{\eta}_{\pm}$ have already been introduced, and the
parameter
$\tilde{\kappa}:=\frac{1}{4}(1-m^2c^4/{E^{'2}_{\tilde{n}_r}})$.
These expressions allow us to study in detail the statistical
nature of the chiral phonon ensemble by means of the Mandel
parameters in Eq.~\eqref{mandel_param}, which have been
represented in figs.~\ref{left_statistics_2D} and
\ref{right_statistics_2D}. In this figures we notice the
following:
\begin{itemize}
\item \textbf{Left-handed regime} $\tilde{\xi}<\xi$.-- In this
limit, the chiral Mandel parameters become:

$Q_l<0\Rightarrow$Sub-Poissonian statistics of quieter left-handed phonons, \\
$Q_r>0\Rightarrow$Super-Poissonian statistics of noisier
right-handed phonons.

\item \textbf{Right-handed regime} $\tilde{\xi}>\xi$.-- In this
limit, the chiral Mandel parameters become:

$Q_l> 0\Rightarrow$Super-Poissonian statistics of noisier left-handed phonons, \\
$Q_r<0\Rightarrow$Sub-Poissonian statistics of quieter
right-handed phonons.

\end{itemize}

Therefore, we observe how the statistical nature of the
relativistic phonons is completely controlled by the relative
coupling strength. Remarkably enough, the level of quantum noise
in the chiral ensembles can be controlled by means of external
parameters. Furthermore, a notorious transition between sub- and
super-Poissonian statistics occurs as the system crosses the
critical region. This fact shows how a quantum phase transition
can also be intrinsically related to the statistical nature of the
system. Note also that close to the critical point $Q_l,Q_r\gg 1$,
and both ensembles become super-Poissonian indicating how quantum
fluctuations diverge at the critical point of a quantum phase
transition.

\subsection{Entanglement at the critical point}

In previous sections, we have seen how several signatures of a
quantum phase transition occur in this relativistic system. We
have described in detail the closure of the energy gap at the
critical point, the existence of an order parameter which turns
out to be the orbital angular momentum, and the divergence of
quantum fluctuations as one explores the critical region. An
additional feature in quantum phase transitions is the divergence
of quantum correlations as the system crosses the critical
region~\cite{osterloh,latorre}. Correlations of a quantum nature,
known as entanglement, lie at the heart of Quantum Mechanics and
are responsible of non-local phenomena in the quantum regime.
Furthermore, they are of  outmost relevance in the fields of
Quantum Information and Computation, where they constitute a
resource for information processing
tasks~\cite{Nielsen_chuang,galindo_martin-delgado,bennet}. In the
particular case of fermions, entanglement has been previously
studied for relativistic field theory~\cite{huerta,calabrese} (see
\cite{vedral} for a recent review). In this section we provide a
description of fermionic entanglement across a quantum phase
transition, that of a relativistic Dirac oscillator subjected to a
magnetic field.

\begin{figure}[!hbp]

\centering

\begin{overpic}[width=4cm]{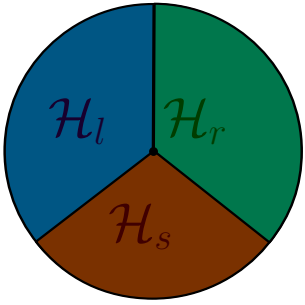}
\put(15,55){\textcolor[rgb]{0.09,0.00,0.18}{\huge{$\mathcal{H}_l$}}}
\put(53,55){\textcolor[rgb]{0.00,0.24,0.00}{\huge{$\mathcal{H}_r$}}}
\put(35,20){\textcolor[rgb]{0.25,0.00,0.00}{\huge{$\mathcal{H}_s$}}}

\end{overpic}
\caption{The relativistic  degrees of freedom can be described as
a tripartite system with a Hilbert space
$\mathcal{H}=\mathcal{H}_r\otimes\mathcal{H}_l\otimes\mathcal{H}_s$
that includes continuous-variables corresponding to the chiral
degrees of freedom, and discrete variables related to the
spinorial degree of freedom.}\label{tripartite_hilbert}

\end{figure}

In our case, the underlying Hilbert space  can be described as a
tripartite system composed of continuous-variables associated to
the chiral degrees of freedom, and discrete-variables related to
the spin degree of freedom (see fig.~\ref{tripartite_hilbert}). To
quantify quantum correlations in this hybrid system we make use of
the machinery of discrete-variable~\cite{plenio} and
continuous-variable~\cite{plenio_cv} entanglement measures. In
particular, we obtain the von Neumann entropy of the reduced
density matrices over all possible bipartitions
\begin{equation}
\begin{split}
&\rho_l:=\text{Tr}_{s}\left(\text{Tr}_r\ket{\psi}\bra{\psi}\right),\\
&\rho_r:=\text{Tr}_{s}\left(\text{Tr}_l\ket{\psi}\bra{\psi}\right),\\
&\rho_s:=\text{Tr}_{l}\left(\text{Tr}_r\ket{\psi}\bra{\psi}\right),
\end{split}
\end{equation}
where the subindexes $l,r,s$ stand for the left-handed,
right-handed, and spinorial degrees of freedom, $\ket{\psi}$ is a
particular pure state, and the von Neumann entropy is defined as
$S(\rho):=-\text{Tr}(\rho \text{log}\rho)$. Such magnitude, when
calculated over reduced density matrices, quantifies the amount of
entanglement between two parties. In our case, it quantifies the
amount of entanglement between different relativistic degrees od
freedom.

\begin{figure}[!hbp]

\centering

\begin{overpic}[width=8.5cm]{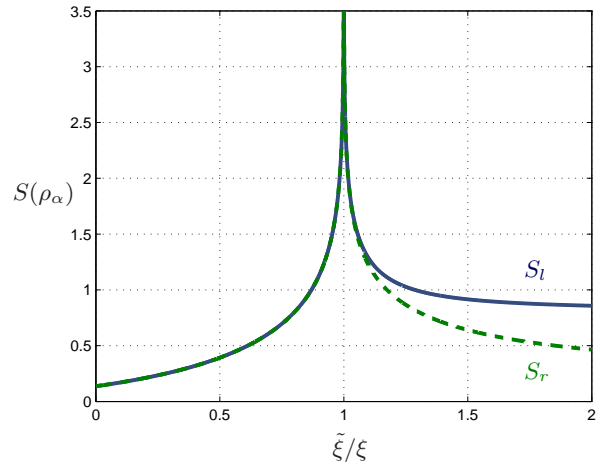}
\put(80,28){\textcolor[rgb]{0.00,0.00,0.37}{$S_l$}}
\put(80,12){\textcolor[rgb]{0.00,0.50,0.00}{$S_r$}}
\put(50,0){$\tilde{\xi}/\xi$} \put(0,40){$S(\rho_{\alpha})$}
\end{overpic}

\caption{Von Neumann entropy of the single-mode reduced  states
$\rho_{\alpha}$, with $\alpha=l,r$ indexing the mode chirality, as
a function of the  coupling strength ratio $\tilde{\xi}/\xi$. One
directly observes that the entanglement of the continuous-variable
degrees of freedom with the rest of the system diverges as one
approaches the critical region.}\label{orb_entanglement_pt}

\end{figure}

\begin{figure}[!hbp]

\centering

\begin{overpic}[width=8.5cm]{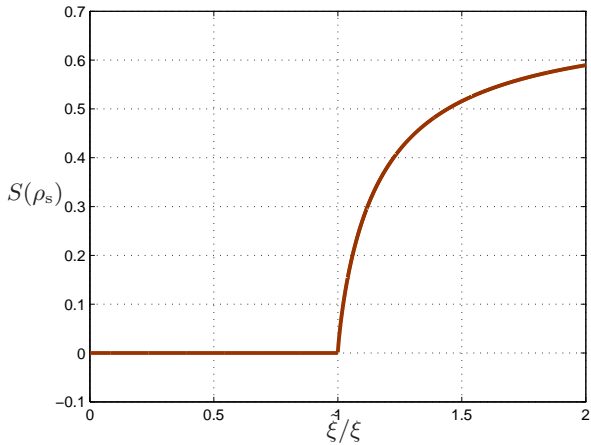}
\put(50,3){$\tilde{\xi}/\xi$} \put(0,40){$S(\rho_{\text{s}})$}
\end{overpic}

\caption{Von Neumann entropy of the spinorial reduced state
$\rho_s$ as a function of  the coupling strength ratio
$\tilde{\xi}/\xi$. We observe that the entanglement of the spin
degree of freedom to the rest of the system increases with the
magnetic field.}\label{spin_entanglement_pt}

\end{figure}

\textbf{Left-handed regime $\tilde{\xi}<\xi$}.-- The ground state
of the relativistic spin-1/2 oscillator in
Eq.~\eqref{left_ground_state} is described by means of a spin-up
two-mode squeezed vacuum state, with the following reduced density
matrices
\begin{equation}
\begin{split}
\rho_l&=\frac{1}{\text{cosh}^2|z|}\sum_{n_l=0}^{\infty}\text{tanh}^{2n_l}|z|\ket{n_l}\bra{n_l},\\
\rho_r&=\frac{1}{\text{cosh}^2|z|}\sum_{n_r=0}^{\infty}\text{tanh}^{2n_r}|z|\ket{n_r}\bra{n_r},\\
\rho_s&=\ket{\chi_{\uparrow}}\bra{\chi_{\uparrow}}.
\end{split}
\end{equation}
The reduced chiral degrees of freedom are described by means of a
mixed thermal state with an effective temperature
$T_{\text{eff}}:=\hbar\omega/2k_B\text{log}(\text{coth}|z|)$,
where $k_B$ is the Boltzmann constant. Remarkably, the entropy of
a thermal state can be given in terms of its mean number of quanta
$\langle n\rangle$~\cite{agarwal,serafini}, and we obtain
$S_l:=S(\rho_l)=\text{sinh}^2|z|\text{log}(1+\text{cosech}^2|z|)+\text{log}(\text{cosh}^2|z|)=S(\rho_r)=:S_r$.
There exists thus a certain amount of chiral entanglement, which
depends on the relative coupling strength and diverges as one
approaches the critical point $(\tilde{\xi}/\xi)_c=1$ ( see
fig.~\ref{orb_entanglement_pt}). Conversely, the reduced spin
density matrix is in a spin-up pure state with vanishing entropy
$S_s:=S(\rho_s)=0$ ( see fig.~\ref{spin_entanglement_pt}). This
remarks the fact that there is no spin-orbit entanglement between
the discrete- and continuous-degrees of freedom, which is in
accordance to the ground state structure in
Eq.~\eqref{left_ground_state}, which is biseparable with respect
to the spin-orbit bipartition.

\textbf{Right-handed regime $\tilde{\xi}>\xi$}.-- In this limit,
the relativistic ground state in Eq.~\eqref{right_ground_state}
can no longer be described by a spin-up squeezed vacuum state, but
rather by means of a SU(1,1)-doublet where chiral coherent states
are entangled with spin states. The reduced density matrices
become
\begin{equation}
\label{right_reduced_states}
\begin{split}
\rho_l&=\frac{1}{\text{cosh}^2|\tilde{z}|}\sum_{\tilde{n}_l=0}^{\infty}\text{tanh}^{2\tilde{n}_l}|\tilde{z}|\left(\gamma_++\frac{\gamma_- \tilde{n}_l}{\text{sinh}^2|\tilde{z}|}\right)\ket{\tilde{n}_l}\bra{\tilde{n}_l},\\
\rho_r&=\frac{1}{\text{cosh}^2|\tilde{z}|}\sum_{\tilde{n}_r=0}^{\infty}\text{tanh}^{2\tilde{n}_r}|\tilde{z}|\left(\gamma_++\frac{\gamma_-(\tilde{n}_r+1)}{\text{cosh}^2|\tilde{z}|}\right)\ket{\tilde{n}_r}\bra{\tilde{n}_r},\\
\rho_s&=\gamma_+\ket{\chi_{\uparrow}}\bra{\chi_{\uparrow}}+\gamma_-\ket{\chi_{\downarrow}}\bra{\chi_{\downarrow}},
\end{split}
\end{equation}
where $\gamma_{\pm}:=(\sqrt{1+2\zeta_r}\pm1)/2\sqrt{1+2\zeta_r}$.
In this regime, the reduced chiral degrees of freedom cannot be
described by means of effective thermal states, nor can the
reduced spin degrees of freedom be identified with a pure state.
Nonetheless, the density matrices in
Eqs.~\eqref{right_reduced_states} are already in diagonal form,
and the von Neumann entropy can be expressed as follows
\begin{equation}
\begin{split}
S_l&=-\sum_{\tilde{n}_l=0}^{\infty}\Theta^{l}_{\tilde{n}_l}\text{log}\Theta^{l}_{\tilde{n}_l},\\
S_r&=-\sum_{\tilde{n}_r=0}^{\infty}\Theta^{r}_{\tilde{n}_r}\text{log}\Theta^{r}_{\tilde{n}_r},\\
S_s&=-\half\left[\text{log}\left(\frac{\zeta_r}{2(1+\zeta_r)}\right)+\frac{1}{\sqrt{1+2\zeta_r}}\text{log}\left(\frac{\sqrt{1+2\zeta_r}+1}{\sqrt{1+2\zeta_r}-1}\right)\right],
\end{split}
\end{equation}
with the parameters
$\Theta^{l}_{\tilde{n}_l}:=\frac{\text{tanh}^{2\tilde{n}_l}|\tilde{z}|}{\text{cosh}^2|\tilde{z}|}\left(\gamma_++\frac{\gamma_-
\tilde{n}_l}{\text{sinh}^2|\tilde{z}|}\right)$, and
$\Theta^{r}_{\tilde{n}_r}:=\frac{\text{tanh}^{2\tilde{n}_r}|\tilde{z}|}{\text{cosh}^2|\tilde{z}|}\left(\gamma_++\frac{\gamma_-
(\tilde{n}_r+1)}{\text{cosh}^2|\tilde{z}|}\right)$. These
magnitudes, which represent entanglement between all possible
bipartitions of the system have been represented in
figs.~\ref{orb_entanglement_pt}~and~\ref{spin_entanglement_pt}. We
observe certain peculiarities with respect to the left-handed
regime, such as a finite spinorial entanglement which tends to its
maximal value $S_s^{\text{max}}=\text{log}2$ as the magnetic field
is raised $\tilde{\xi}/\xi\to\infty$. We also observe that the
continuous-variable entanglement is no longer equal for opposed
chiral modes, which is a consequence of a finite orbital angular
momentum in the ground state. In addition, one can see from
Eq.~\eqref{right_ground_state} that the relativistic ground state
presents genuine tripartite entanglement, where all the degrees of
freedom are correlated among each other.

\section{Conclusions}

In this work we have studied the properties of a relativistic
spin-1/2 Dirac oscillator subjected to a constant magnetic field.
The relativistic Hamiltonian can be mapped onto a couple
Jaynes-Cummings and Anti-Jaynes-Cummings terms, which describe the
interaction between the relativistic spinor and bosons that carry
an intrinsic chirality. These models, which are of paramount
importance in Quantum Optics, become  useful in this relativistic
scenario, and  allow an insightful description of interplay
between opposed chirality interactions. The study of this
relativistic system in the limits of a weak and strong magnetic
field points towards the existence of two different phases, each
described by an opposed chirality. In the intermediate regime, an
intriguing trade-off between left- and right-handed interactions
leads to the appearance of a quantum phase transition, which we
have described in full detail.

The remarkable possibility to find the exact solution of this
relativistic system has allowed us to describe analytically
several properties of this phase transition. We have discussed the
non-analytic nature of the energy spectrum at the critical point,
and the sudden extinction of the energy gap, which are
archetypical properties of quantum phase transitions. We have also
proved that there is an order parameter, the $z$-component of the
orbital angular momentum, that witnesses the chiral phase change.
In the critical region, we have shown that the quantum
fluctuations in the fermion position diverge, which is a typical
sign of quantum phase transition. Conversely, the quantum
fluctuations in momentum vanish at the critical point, since a
particular squeezing of this canonical variable occurs due to the
interplay of the interactions. We have discussed the bosonic
ensemble statistics, and their behavior across the phase
transition. In this regard, we have found that super- or
sub-Poissonian nature of such ensemble can be controlled at will
by an appropriate modification of the magnetic field strength.

Finally, the entanglement properties between the various degrees
of freedom of this relativistic system has also been accomplished.
This hybrid system consists of a couple of continuous-variable
degrees of freedom associated to the chiral modes, and a
discrete-variable associated to the spin, therefore constituting a
tripartite system. We have shown that an entanglement measure, the
von Neumann entropy of the reduced states, over all possible
bipartitions can be obtained for the system ground state. At the
critical point, the quantum correlations of the chiral degrees of
freedom diverge, which can be also considered as a sign of quantum
phase transitions.

\vspace{2ex}

\noindent\emph{Acknowledgements}.-- We  acknowledge financial
support from the Spanish MEC project FIS2006-04885, the project
CAM-UCM/910758 (AB and MAMD) and the UCM project PR1-A/07-15378
(AL). Additionally, we acknowledge support from a FPU MEC grant
(AB), and the ESF Science Programme INSTANS 2005-2010 (MAMD).

\appendix

\section{SU(1,1) coherent states}
\label{apendix_a}

In this appendix we review the properties of two-mode SU(1,1)
coherent states. Such states arose from a group theoretical
approach to the generalization of the usual Glaubler coherent
states~\cite{glauber}, which can be constructed for an arbitrary
Lie group~\cite{perelomov}. In this case, we shall restrict our
attention to the SU(1,1) group, whose generators $\{K_0,K_+,K_-\}$
have the following Lie Algebra
\begin{equation}
\label{su11_lie_algebra} [K_0,K_{\pm}]=\pm K_{\pm}, \hspace{2ex}
[K_-,K_+]=2K_0.
\end{equation}
The Schwinger representation of is
algebra~\eqref{su11_lie_algebra} consists on a pair of commuting
bosonic harmonic oscillator operators $a_r,a_l$ related to the
group generators as
follows~\cite{two_mode_coherent_gerry,two_mode_coherent_knight}
\begin{equation}
K_0=\half\left(a_r^{\dagger}a_r+a_l^{\dagger}a_l+1\right),\hspace{1ex}
K_+=a_r^{\dagger}a_l^{\dagger}=(K_-)^{\dagger}.
\end{equation}
This two-mode representation is very useful, since one can use the
disentangling theorem~\cite{perelomov} to calculate the SU(1,1)
coherent states
$\ket{z,n_l}=U_{\alpha}^{\dagger}\ket{n_l}\ket{\text{vac}}_r$ in
Eq.~\eqref{su11_left_coherent_states} in the usual Fock basis
$\ket{n_r,n_l}$
\begin{equation}
\label{exp_disentangling}
U_{\alpha}^{\dagger}=\ee^{z(K_+-K_-)}=\ee^{\text{tanh}zK_+}\ee^{\text{log}(\text{cosh}^{-2}z)K_0/2}\ee^{-\text{tanh}zK_-}.
\end{equation}
Such factorization of the exponential~\eqref{exp_disentangling}
allows the following expression for the SU(1,1) coherent states
\begin{equation}
\ket{z,n_l}=\mathcal{N}_{n_l}\sum_{m=0}^{\infty}\sqrt{\frac{(m+n_l)!}{n_l!m!}}(-1)^m\text{tanh}^m|z|
\ket{m+n_l,m},
\end{equation}
which has been used in Eq.~\eqref{su11_left_coherent_states} of
the left-handed regime $\xi>\tilde{\xi}$. Analogously, one can
obtain the corresponding SU(1,1) coherent states in the
right-handed regime, using a similar disentangling theorem
\begin{equation}
\ket{\tilde{z},\tilde{n}_r}=\mathcal{N}_{\tilde{n}_r}\sum_{\tilde{m}=0}^{\infty}\sqrt{\frac{(\tilde{m}+\tilde{n}_r)!}{\tilde{n}_r!\tilde{m}!}}(-1)^{\tilde{m}}\text{tanh}^{\tilde{m}}|\tilde{z}|
\ket{\tilde{m}+\tilde{n}_r,\tilde{m}},
\end{equation}
which was used in Eq.~\eqref{right_coh_states}.

In the Lie algebraic language, the Casimir operator $C$ commutes
with the generators $\{K_0,K_{\pm}\}$ of the algebra. In the
SU(1,1) case, the Casimir operator turns out to be the
$z$-component of the orbital angular momentum $C=L_z$, which is
therefore left unchanged under the action of the squeezing
operator $[U_{\alpha}^{\dagger},L_z]=0$. This fact has been used
in section~\ref{sectionV}, where the expected value of such
Casimir operator is used as an order parameter that reveals the
existence of a quantum phase transition.

SU(1,1) coherent states present several remarkable non-classical
properties, which attain an special interest in Quantum Optics
since they can be engineered in experiments by means of a
degenerate parametric amplifier. In the relativistic setting, they
arise naturally as the eigenstates of a relativistic oscillator
subjected to a magnetic field along the $z$-axis. In particular,
we shall be interested in the following properties of such states,
which have been extensively studied
in~\cite{two_mode_coherent_gerry},
\begin{equation}
\label{left_coh_exp_values}
\begin{split}
&\langle n_r
\rangle_{|z,n_l\rangle}=(n_l+1)\text{sh}^2|z|,\\
&\langle n_l
\rangle_{|z,n_l\rangle}=n_l\text{ch}^2|z|+ \text{sh}^2|z|,\\
&\langle n_r^2
\rangle_{|z,n_l\rangle}=(n_l+1)^2\text{sh}^4|z|+(1+n_l)\text{ch}^2|z|\text{sh}^2|z|,\\
&\langle n_l^2
\rangle_{|z,n_l\rangle}=n_l^2\text{ch}^4|z|+\text{sh}^4|z|+(1+3n_l)\text{ch}^2|z|\text{sh}^2|z|,\\
\end{split}
\end{equation}
for the SU(1,1) coherent states in the left-handed regime, where
$\text{sh}|z|:=\text{sinh}|z|$,and $\text{ch}|z|:=\text{cosh}|z|$,
and we use the chiral number operators $n_r:=a^{\dagger}_ra_r$,
and $n_l:=a^{\dagger}_la_l$ . Analogously, the corresponding
coherent states in the right-handed regime fulfill the following
\begin{equation}
\label{right_coh_exp_values}
\begin{split}
&\langle \tilde{n}_l
\rangle_{|\tilde{z},\tilde{n}_r\rangle}=(1+\tilde{n}_r)\text{sh}^2|\tilde{z}|,\\
&\langle \tilde{n}_r
\rangle_{|\tilde{z},\tilde{n}_r\rangle}=\tilde{n}_r\text{ch}^2|\tilde{z}|+\text{sh}^2|\tilde{z}|,\\
&\langle \tilde{n}_l^2
\rangle_{|\tilde{z},\tilde{n}_r\rangle}=(\tilde{n}_r+1)^2\text{sh}^4|\tilde{z}|+(1+\tilde{n}_r)\text{ch}^2|\tilde{z}|\text{sh}^2|\tilde{z}|,\\
&\langle \tilde{n}_r^2
\rangle_{|\tilde{z},\tilde{n}_r\rangle}=\tilde{n}_r^2\text{ch}^4|\tilde{z}|+\text{sh}^4|\tilde{z}|+(1+3\tilde{n}_r)\text{ch}^2|\tilde{z}|\text{sh}^2|\tilde{z}|,\\
\end{split}
\end{equation}
where $\text{sh}|\tilde{z}|:=\text{sinh}|\tilde{z}|$,and
$\text{ch}\tilde{r}:=\text{cosh}|\tilde{z}|$, and we use the
chiral number operators
$\tilde{n}_r:=\tilde{a}^{\dagger}_r\tilde{a}_r$, and
$\tilde{n}_l:=\tilde{a}^{\dagger}_l\tilde{a}_l$. These
expressions~\eqref{left_coh_exp_values},\eqref{right_coh_exp_values}
are extremely useful for the study of the relativistic phonon
statistics in Section~\ref{sectionV}.


\begin{references}
\bibitem{notaI}
Note that these creation-annihilation operators do not represent
normal modes since they do not commute with each other, but rather
 fulfill $[a_i,\tilde{a}_j]=\delta_{ij}\mu$,
$[a_i,\tilde{a}^{\dagger}_j]=\delta_{ij}\tilde{\mu}$, where
$\mu=:\frac{1}{2}\left(\frac{\Delta}{\tilde{\Delta}}-\frac{\tilde{\Delta}}{\Delta}\right)$,
and
$\tilde{\mu}=\frac{1}{2}\left(\frac{\Delta}{\tilde{\Delta}}+\frac{\tilde{\Delta}}{\Delta}\right)$.
\end{references}
\end{document}